# Properties of MgB$_2$ Films Grown at Various Temperatures by Hybrid Physical-Chemical Vapour Deposition

**Ke Chen[1], Menno Veldhorst[1,2], Che-Hui Lee[3], Daniel R. Lamborn[3], Raymond DeFrain[3], Joan M. Redwing[3], Qi Li[1], and X. X. Xi[1,3]**

1.  Department of Physics, The Pennsylvania State University, University Park, Pennsylvania 16802, USA

2.  The Faculty of Science and Technology and MESA+ Institute for Nanotechnology, University of Twente, The Netherlands

3.  Department of Materials Science and Engineering, The Pennsylvania State University, University Park, Pennsylvania 16802, USA

## Abstract

A Hybrid Physical-Chemical Vapour Deposition (HPCVD) system consisting of separately controlled Mg-source heater and substrate heater is used to grow MgB$_2$ thin films and thick films at various temperatures. We are able to grow superconducting MgB$_2$ thin films at temperatures as low as 350°C with a $T_{c0}$ of 35.5 K. MgB$_2$ films up to 4 μm in thickness grown at 550°C have $J_c$ over $10^6$ A/cm$^2$ at 5 K and zero applied field. The low deposition temperature of MgB$_2$ films is desirable for all-MgB$_2$ tunnel junctions and MgB$_2$ thick films are important for applications in coated conductors.

## 1. Introduction

Since the discovery of 39 K superconductivity in MgB$_2$ in 2001 [1], there have been efforts to fabricate MgB$_2$ Josephson junctions to replace Nb-based junctions and take advantage of the much higher $T_c$ than Nb. The simpler structure and longer coherence length than cuprates also make MgB$_2$ attractive. MgB$_2$ films have been grown using *ex situ* annealing B film in Mg vapour [2], pulse laser deposition (PLD) [3], sputtering [4], molecular beam epitaxy (MBE) [5] and co-evaporation [6]. Due to the low sticking coefficient of Mg at high temperature, MgB$_2$ films are limited to grow at 300 °C or lower by most *in situ* physical vapour deposition methods [7], which is much lower than the optimal temperature of 1080 °C, which is about half of the MgB$_2$ melting point [7]. To overcome this difficulty, Hybrid Physical-Chemical Vapour Deposition (HPCVD) was developed to grow high-quality MgB$_2$ films at about 700 °C under high Mg vapour pressure produced by thermally evaporated Mg chips with B$_2$H$_6$ as the B source [8] in a high purity H$_2$ atmosphere. HPCVD MgB$_2$ films exhibit superior properties such as high critical temperature ($T_c$), excellent crystallinity, and low impurity and defect scattering [9]. Excellent tunnelling characteristics have been observed on Josephson tunnel junctions made on HPCVD films with Pb counter-electrode and natural oxide tunnel barriers [10, 11]. These facts imply a promising all-MgB$_2$ Josephson tunnel junction if both electrodes are made by HPCVD. However, despite the existence of several reports on all-MgB$_2$ tunnel junctions made by MBE [12] and sputtering [13], an all-MgB$_2$ tunnel junction has not yet successfully been made using HPCVD films. Most of our attempts to make all-MgB$_2$ tunnel junctions *in situ* or *ex situ* using HPCVD have failed due to random shorts inside the tunnel barrier. The failure of the natural oxide barrier may be attributed to the relatively high deposition temperature, around 700 °C used in the original HPCVD system during the top electrode deposition. A low deposition temperature for the top electrode is desirable for preventing this failure. In addition, a low temperature deposition process is also desired for fabricating multilayer devices and for deposition on materials such as silicon, copper, and plastics, which alloy with magnesium or melt at high temperature.

The original HPCVD system at Penn State [8] uses a single susceptor to heat both substrate and Mg chips, therefore the Mg partial pressure is linked to the substrate temperature. Although this works well for high temperature deposition around 700 °C, it is not possible to grow good quality films at temperatures lower than 650 °C. In order to reduce the substrate temperature, it is necessary to independently control the Mg vapour pressure to satisfy the growth conditions at a particular growth temperature [14, 15], determined by thermodynamics [16] and kinetics [17]. In this paper, we report the development of an HPCVD system with separately controlled Mg-source heater and substrate heater that enables the deposition of superconducting MgB$_2$ films from 350 °C to 750 °C. In addition to flexible deposition temperatures, the large Mg source volume of this system allows long deposition time to grow thick MgB$_2$ films. Structural and transport properties of MgB$_2$ films deposited at different temperatures and with different thicknesses are measured and discussed.

## 2.  Experimental details

Figure 1 shows (a) the picture and (b) the schematic of the HPCVD system used in this work. It consists of a substrate heater (3) and an Mg-source heater (6), which are separately heated by resistive heating elements (1, 7) and monitored by thermocouples (2, 9). A molybdenum crucible (4) about 2.2 cm in diameter and 2.5 cm deep is used for the Mg source, which is able to contain 10 g of Mg ingots (5). A stainless steel tube with 5 mm inner diameter (10) is used to introduce the ultra-high-purity (UHP) H$_2$ carrier gas and the 5% B$_2$H$_6$ in H$_2$ precursor gas for boron into the chamber for the deposition. The end of the tubing points to the substrate so that the gases are directed towards the substrate. Most of the tube is shielded by three layers of stainless steel tubing (11) to reduce thermal radiation from the substrate and Mg-source heaters and prevent the premature decomposition of B$_2$H$_6$. A quartz tube (8) of 13 cm in diameter is employed that encloses both the substrate and Mg-source heaters and the gas tubing. It regulates the gas flow around the Mg source and substrate and collects unreacted Mg vapour for efficient and safe cleaning. The Mg-source heater sits on a stage (12) that can move up and down to adjust the distance between the Mg source and the substrate. The substrate is held by a substrate holder on a rotation stage (13) below the heating element and heated by radiation. The rotation of the substrate helps to achieve uniformity of the film.

The MgB$_2$ films were deposited at a total pressure of 17 ~ 35 Torr, total gas flow rate of 500 sccm and B$_2$H$_6$ flow rate of 0.3 ~ 15 sccm. The Mg source temperature ranged from 621 ~ 700 °C depending on the growth temperature which ranged from 350 ~ 750 °C. The total pressure was measured by a capacitance manometer and was controlled by a throttle valve. A summary of the growth conditions is shown in Table 1. Under the conditions of growth temperature at 550 °C and Mg-source temperature at 700 °C, the amount of Mg in the Mo crucible is sufficient for a deposition longer than 90 minutes. SiC (0001) was used as the substrate in this work. Similar results were also obtained for *c*-cut sapphire and MgO (111) substrates.

The substrate temperature during the deposition was monitored by the substrate heater thermocouple, which is fixed at the back of a boron nitride holder that holds the Mo heating element. To correlate the substrate temperature to the reading of the thermocouple, a calibration was made in which a thermal couple was attached to the substrate surface and its reading taken as the actual substrate temperature. The actual substrate temperature was 100 ~ 150 °C lower than the substrate heater thermocouple reading. All the substrate temperatures reported in this work are estimated using this calibration based on the readings of the substrate heater thermocouple.

### 3.   Results and Discussion

The lowest deposition temperature we were able to achieve using the HPCVD system with separately controlled Mg-source heater and substrate heater was 350 °C. This temperature is close to the limit for B$_2$H$_6$ decomposition by pyrolytic reaction at a reasonable rate, which was confirmed by deposition of a pure boron film using B$_2$H$_6$. Table 1 shows the growth conditions for the films reported in this paper. The zero-resistivity critical temperature ($T_{c0}$), transition temperature width ($\Delta T_c$), residual-resistivity-ratio (*RRR*), resistivity at 42 K [$\rho$(42 K)], 300 K [$\rho$(300 K)], and the resistivity difference [$\Delta\rho=\rho$(300 K)-$\rho$(42 K)] of the films are listed in Table 2. The *RRR* is defined as *RRR* = $\rho$(300 K)/$\rho$(42 K), which is commonly used as an indication of the extent of scattering in the film. Larger *RRR* reflects fewer scatterings in the film, therefore is more desired for a pure clean film.

The structural properties of MgB$_2$ films deposited at different temperatures was studied by x-ray diffraction (XRD). Figure 3 shows $\theta$–$2\theta$ scans for two films deposited at 620 °C and 350 °C on a SiC (0001) substrate, respectively. The XRD spectrum for the film deposited at 620 °C shows strong MgB$_2$ (0001) and (0002) peaks, and no other reflection peaks from MgB$_2$ were observed. It suggests that the MgB$_2$ film is *c*-axis oriented. On the other hand, the XRD spectrum for the film deposited at 350 °C shows weak MgB$_2$ (10$\bar{1}$1), (0002), (11$\bar{2}$0), and (11$\bar{2}$1) peaks, suggesting that it is polycrystalline. It also shows relatively sharp MgO peaks which we believe are from MgO impurities inside the film. A broad Mg(OH)$_2$ (0001) peak appears in both spectra, which may be due to the degradation of the surface of the MgB$_2$ films in ambient environment during the XRD measurement which usually took more than 3 hours [18]. The film surface morphology was studied by atomic force microscopy (AFM). Figure 2 shows AFM images for the films deposited at (a) 350 °C and (b) 620 °C. The film deposited at 620 °C is smooth, with a root-mean-square (rms) roughness of 4 nm. However, the film deposited at 350 °C is rough, with an rms roughness of 300 nm. The morphology is consistent with the polycrystalline structure of the film, and the grain size is ~ 500 nm.

Figure 4 shows the resistivity-temperature characteristics of thin MgB$_2$ films grown at 350, 380, 400, 500, 620, and 750 °C. The film thicknesses range from 100 nm to 200 nm as listed in Table 1. $T_{c0}$ as a function of deposition temperature is plotted in Figure 5. Among all the films, the film deposited at 620 °C has the highest $T_{c0}$ of 40.8 K, the largest *RRR*, and the smallest $\rho$(42 K). It is consistent with the results from AFM and XRD, which show that the film deposited at 620 °C has better crystallinity than the films deposited at lower temperatures. As the substrate temperature decreases, $T_{c0}$ in general decreases and resistivity increases. We can see that the film deposited at 350 °C has a $T_{c0}$ of 35.5 K and a residual resistivity of 88 µΩcm. The $T_{c0}$ is comparable to or a little higher than films deposited at 300 °C by other techniques including MBE [19] and sputtering [20]. The film deposited at the highest temperature 750 °C has poorer properties than the film deposited at 620 °C, which may be due to the premature decomposition of B$_2$H$_6$ in the gas delivery tube. When the substrate temperature was at 750 °C, the temperature at the tip of the gas delivery tube reached over 300 °C, and a layer of brownish deposit, presumably boron, was found inside the tube after the deposition. Due to the much larger hot surface areas and the lack of an effective purging procedure before deposition, contaminants from both the heaters and the gas delivery tubing may also become a problem causing lower *RRR* and higher $\rho$(42 K) than films from the original HPCVD system. We also observe the general trend that films deposited at higher temperature up to 620 °C have a lower $\rho$(42 K) due to better crystallinity and therefore less scattering by defects at higher deposition temperature. One exception is the 150 nm-thick film deposited at 400 °C which has higher $\rho$(42 K) and lower $T_{c0}$ than the 200 nm-thick film deposited at 380 °C possibly due to more scattering from the surface of the thinner film. As Rowell [21] has pointed out, $\Delta\rho=\rho$(300 K)-$\rho$(42 K) is an indicator of connectivity of grains in an MgB$_2$ sample. A well-connected MgB$_2$ sample should have a $\Delta\rho$ around 8 µΩcm. Among the thin films deposited at different temperatures, only the one deposited at 620 °C has a $\Delta\rho$ close to this value. Any deviation from this optimum temperature leads to a larger $\Delta\rho$ of about 32 $\mu\Omega$cm, which does not change further with temperature. This value, while larger than in the films deposited at optimum temperature, is relatively small compared to MgB$_2$ samples made by many other techniques, indicating a limited reduction in connectivity in the films deposited at various temperatures.

Figure 6 shows the resistivity versus temperature dependence of MgB$_2$ films with thicknesses of 1.6, 3.0 and 4.0 $\mu$m, all deposited at the substrate temperature of 550°C. Among these films, the 1.6 $\mu$m-thick film has the highest $T_{c0}$ of 40.4 K, the largest *RRR* of about 7, and the lowest $\rho$(42 K) of about 2 $\mu\Omega$cm. As the film thickness increases to 4.0 $\mu$m, $T_{c0}$ decreases to 39.7 K. Since the higher-than-bulk $T_{c0}$ in HPCVD thin films is due to the epitaxial strain in the films [22], the slight decrease in $T_{c0}$ may be partly due to the diminishing epitaxial strain in thick films. The $\Delta\rho$ increases as the film thickness increases, indicating the grains in thicker films are less connected than those in thinner films. Figure 7 shows the scanning electron microscopy (SEM) images of the surface [(a) and (c)] and cross-section [(b) and (d)] of the 1.6 $\mu$m- and 3.0 $\mu$m- thick films, respectively. Regularly-oriented grains in (a) and (b) suggest that the 1.6 $\mu$m-thick film is possibly *c*-axis oriented, similar to other *c*-axis oriented thin films grown by the original HPCVD system [8]. The randomly-oriented grains in (c) and (d) suggest that the 3.0 $\mu$m-thick film is polycrystalline with grain size about 3 $\mu$m.

We have carried out magnetization (*M*) measurements and determined the critical current density $J_c$ from the *M-H* loop using the Bean model [23]. The $J_c$ versus magnetic field *B* dependences of three films of 0.15, 1.6, and 4.0 $\mu$m in thickness are plotted in Figure 8. The solid symbols are the results for $T = 5$ K and the open symbols are for $T = 20$ K. At zero field and $T = 5$ K, all films exhibit $J_c$ of about $1 \sim 2 \times 10^6$ A/cm$^2$, which is comparable to films made with other methods. Thicker films exhibit lower $J_c$ than thinner films, which may be because grains are less connected in thicker films, as indicated by larger $\Delta\rho$ in Table 2. In addition, the $J_c$'s of thicker films decrease faster than those of thinner films as $B$ increases. The faster decrease in $J_c$ signals weaker vortex pinning in thicker films; the mechanism is currently under investigation.

## 4.  Conclusion

In summary, we have deposited MgB$_2$ thin films at different substrate temperatures using an HPCVD system with separate substrate heater and Mg-source heater and successfully extended the deposition temperature for MgB$_2$ films to as low as 350 °C. Films with thickness up to 4 μm have been grown. The MgB$_2$ film grown at 350 °C is polycrystalline and exhibits a $T_{c0}$ of 35.5 K. It has the potential to be used as the top electrode for all-MgB$_2$ tunnel junctions. The 4 μm-thick MgB$_2$ films grown at 550 °C has a $J_c > 10^6$A/cm$^2$ at 5 K and zero applied magnetic field.

## Acknowledgments

This work is supported in part by ONR under Grant Nos. N00014-07-1-0079 (X.X.X) and N00014-06-1-1019 (J.M.R) and by NSF under Grant No. DMR-0405502 (Q.L.). We would like to acknowledge the assistance of Alexej Pogrebnyakov, Shufang Wang, and Wenqing Dai during the setup of the integrated HPCVD system. The authors acknowledge use of facilities at the PSU Site of the NSF NNIN under Agreement # 0335765.

**Figure Captions**

Figure 1. (a) The picture and (b) the schematic of the HPCVD system. The labels in (b) are: 1 – molybdenum wire, 2 – thermocouple, 3 – substrate heater, 4 – molybdenum crucible, 5 – Mg ingots, 6 – Mg-source heater, 7 – molybdenum wire, 8 – quartz tube, 9 – thermocouple, 10 – stainless steel tubing for gas delivery, 11 – stainless steel radiation shields, 12 – Mg-source heater stage, 13 – substrate holder stage.

Figure 2. AFM images of the surface of MgB$_2$ films grown at (a) 350 °C and (b) 620 °C. The image area is 5 μm by 5 μm. The height scales are (a) 2 μm and (b) 30 nm.

Figure 3. XRD $\theta$–$2\theta$ scans of MgB$_2$ films grown at 350 °C and 620 °C.

Figure 4. The resistivity versus temperature dependences of MgB$_2$ thin films grown at various temperatures. (a) shows the full temperature range; (b) shows the temperature region near $T_c$.

Figure 5. The $T_{c0}$ versus deposition temperature dependence of all the films reported in this work.

Figure 6. The resistivity versus temperature dependences of MgB$_2$ thick films of various thickness.

Figure 7. SEM images of (a)(b) 1.6 μm-thick and (c)(d) 4.0 μm-thick MgB$_2$ films. (a)(c) are the surface images and (b)(d) are the cross-section images. The scale bar in each image is 2 μm long.

Figure 8. The critical current versus applied magnetic field dependences of MgB$_2$ films of various thicknesses. Solid (open) symbols are the results for $T = 5$ K (20 K).

**Table 1.** Growth conditions for the films reported in this paper.

| $T_{sub}$ (°C) | Film thickness (nm) | B$_2$H$_6$ flow rate[a] (sccm) | H$_2$ flow rate (sccm) | $P_{total}$ (Torr) | $T_{Mg}$ (°C) | Growth time (min) |
|---|---|---|---|---|---|---|
| 350 | 200 | 0.3 | 500 | 30 | 621 | 25 |
| 380 | 200 | 0.3 | 500 | 30 | 632 | 20 |
| 400 | 150 | 0.4 | 500 | 30 | 690 | 20 |
| 500 | 100 | 0.4 | 500 | 30 | 705 | 40 |
| 620 | 100 | 2 | 498 | 30 | 795 | 10 |
| 750 | 100 | 10 | 490 | 35 | 840 | 5 |
| | | | | | | |
| 550 | 1600 | 10 | 490 | 17 | 685 | 30 |
| 550 | 3000 | 14 | 486 | 17 | 700 | 40 |
| 550 | 4000 | 15 | 485 | 17 | 700 | 90 |

[a] 5% B$_2$H$_6$ + 95% H$_2$ gas mixture was used as the boron source in all film depositions. The flow rate of the gas mixture is listed here.

**Table 2.** Characteristics of the MgB$_2$ films reported in this paper.

| Growth Temp. (°C) | Film thickness (nm) | $T_{c0}$ (K) | $\Delta T_c$ (K) | *RRR* | $\rho$(42 K) (μΩcm) | $\rho$(300 K) (μΩcm) | $\Delta\rho$ (μΩcm) |
|---|---|---|---|---|---|---|---|
| 350 | 200 | 35.5 | 0.9 | 1.4 | 88 | 120 | 32 |
| 380 | 200 | 36.5 | 0.6 | 1.4 | 61 | 96 | 35 |
| 400 | 150 | 35.9 | 0.8 | 1.4 | 78 | 109 | 31 |
| 500 | 100 | 37.7 | 0.7 | 1.7 | 47 | 79 | 32 |
| 620 | 100 | 40.8 | 0.1 | 5.9 | 2.2 | 13 | 11 |
| 750 | 100 | 38.7 | 0.5 | 3.6 | 12 | 43 | 31 |
| | | | | | | | |
| 550 | 1600 | 40.4 | 0.4 | 7.0 | 2.0 | 13.9 | 12 |
| 550 | 3000 | 40.1 | 0.3 | 6.0 | 3.6 | 21.6 | 18 |
| 550 | 4000 | 39.7 | 0.5 | 5.6 | 6.4 | 36 | 30 |

Fig. 1

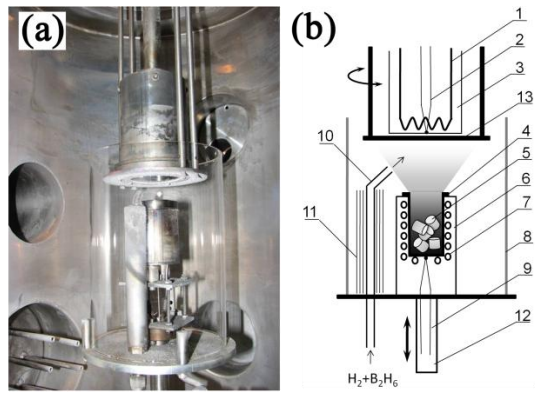

Fig. 2

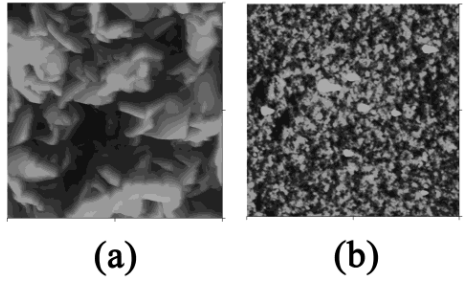

**(a)**    **(b)**

Fig. 3

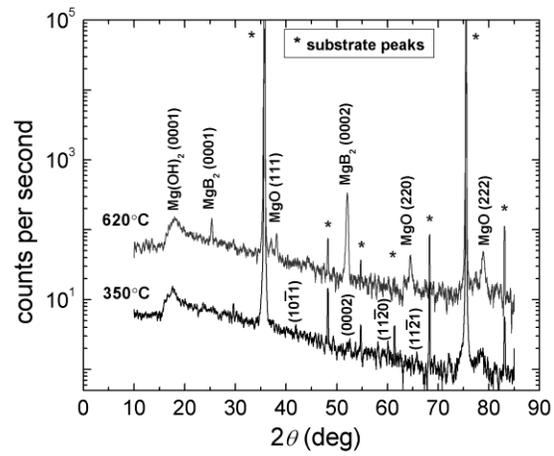

Fig. 4

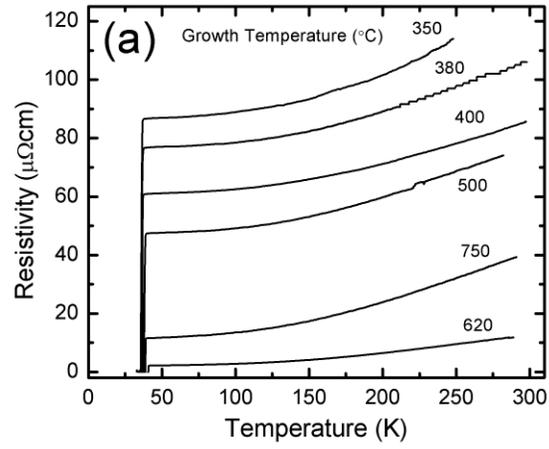

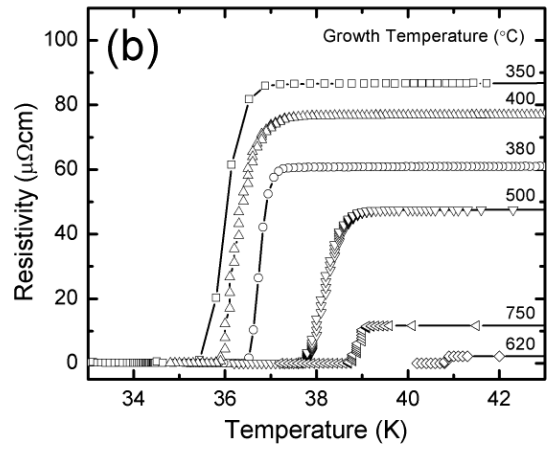

Fig. 5

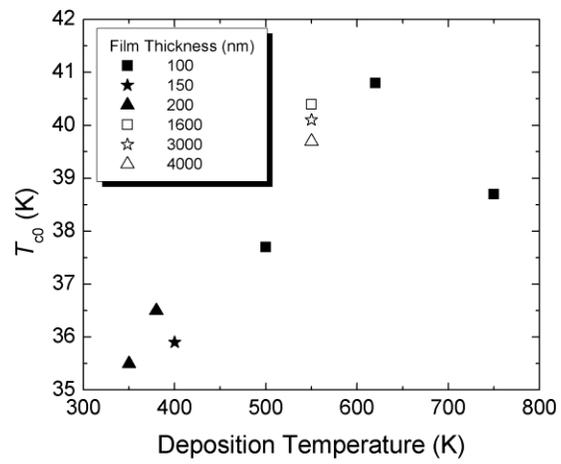

Fig. 6

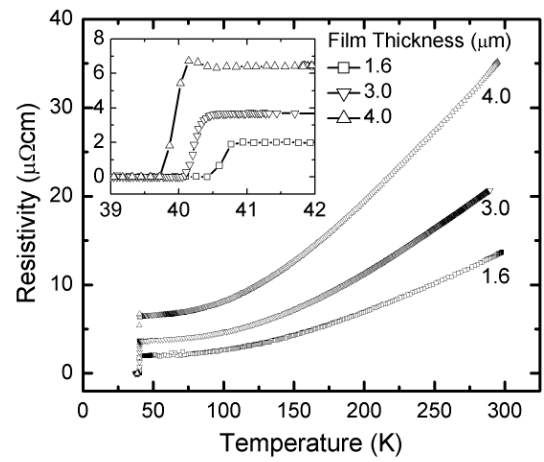

Fig. 7

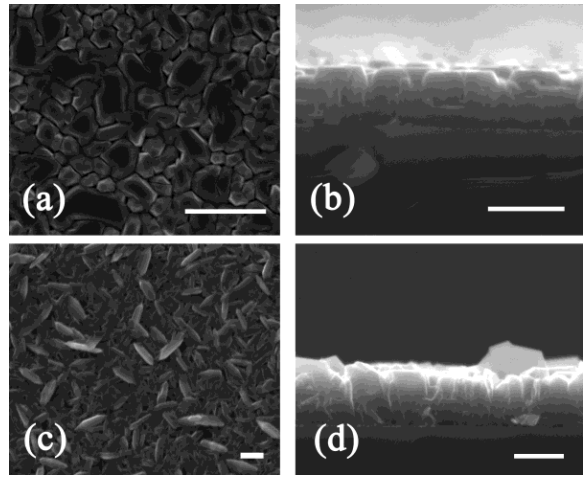

Fig. 8

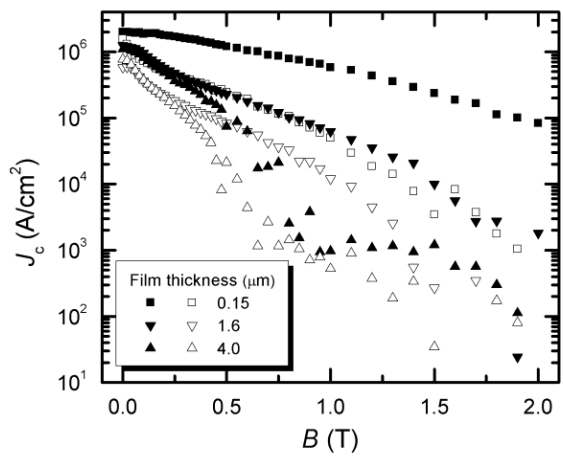